\documentclass[journal]{IEEEtran}
\usepackage{amsmath,amsfonts}
\usepackage{array}
\usepackage[caption=false,font=normalsize,labelfont=sf,textfont=sf]{subfig}
\usepackage{textcomp}
\usepackage{stfloats}
\usepackage{url}
\usepackage{verbatim}
\usepackage{subfig}
\usepackage{graphicx}
\usepackage{cite}
\usepackage{xcolor}
\usepackage{enumitem}
\begin{document}

\setlength{\abovecaptionskip}{0pt}
\setlength{\textfloatsep}{2pt}

\title{Semantics-Aware Unified Terrestrial Non-Terrestrial 6G Networks}

\author{Erfan Delfani, Agapi Mesodiakaki, Leandros Tassiulas, and Nikolaos Pappas \\
\thanks{Erfan Delfani and Nikolaos Pappas are with the Department of Computer and Information Science, Linköping University, Sweden, email: \{\texttt{erfan.delfani, nikolaos.pappas\}@liu.se}. \\ 
Agapi Mesodiakaki is with Aristotle University of Thessaloniki, and Center for Interdisciplinary Research and Innovation, Thessaloniki, Greece, email: \texttt{amesodia@csd.auth.gr}. \\
Leandros Tassiulas is with Yale University, New Haven, CT, USA, email: \texttt{leandros.tassiulas@yale.edu}. \\
This work has been supported in part by the Swedish Research Council
(VR), ELLIIT, and the European Union (ELIXIRION, 101120135, 6G-LEADER, 101192080, and ETHER, 101096526).}}

\maketitle

\begin{abstract}
The integration of Terrestrial and Non-Terrestrial Networks (TN-NTNs), introduced in 5G, is advancing toward a unified and seamless \emph{network of networks} in Sixth-Generation (6G). This evolution markedly increases the volume of generated and exchanged data, imposing stringent technical and operational requirements along with higher cost and energy consumption. Consequently, efficient management of data generation and transmission within this unified architecture has become essential. In this article, we investigate semantics-aware information handling in unified TN-NTNs, where data communication between distant TN nodes is enabled via an NTN. We consider an Internet of Things (IoT) monitoring system in which status updates from a remote Energy Harvesting (EH) device are delivered to a destination monitor through a network of Low Earth Orbit (LEO) satellites. We leverage semantic metrics, such as Query Version Age of Information, which collectively capture the timeliness, relevance, and utility of information. This approach minimizes the transmission of stale, uninformative, or unusable information, thereby reducing the volume of data that must be transmitted and processed. The result is a substantial reduction in energy consumption and data exchange within the network—achieving up to $73\%$ lower energy-charging requirements and fewer transmission demands than the state of the art—without compromising the conveyed information.
\end{abstract}


\section{Introduction}

Integrating Non-Terrestrial Networks (NTNs) with Terrestrial Networks (TNs) promises extended, seamless coverage and ubiquitous connectivity on a global scale. This integration establishes a unified communication infrastructure that enables efficient exchange of information across diverse environments, including ships, airplanes, and remote rural areas\cite{geraci2022integrating}. Operating within a three-dimensional (3D) architecture that spans terrestrial, aerial, and space layers\cite{ntontinether}, the resulting network supports a wide range of applications, such as global asset monitoring, smart agriculture, autonomous vehicles, environmental monitoring, industrial IoT, Public Protection and Disaster Relief (PPDR), and remote healthcare\cite{ntontin2025vision}.

Nevertheless, as this integration advances toward unified 6G 3D networks, the number of energy-consuming network elements and the associated compute and storage requirements increase significantly due to the growth of generated and transmitted data. Remote IoT devices, aerial nodes (e.g., Unmanned Aerial Vehicles (UAVs)), and space nodes are typically battery-powered with limited or no access to energy supplies, underscoring the necessity of energy-efficient operation under strict resource constraints to maintain reliable services \cite{jia2025nfv}. Moreover, bandwidth and propagation limitations in NTNs can lead to congestion or delays, highlighting the importance of effective resource allocation and prioritization of critical communications (e.g., emergency services) during peak demand. Efficient traffic management is therefore essential to ensure timely communication between the TN and NTN layers without compromising Quality of Service (QoS).

In this context, the \emph{semantics-aware communication} paradigm employs goal-oriented approaches to generate and transmit the most informative data, considering the innate and task-specific attributes of information—including \emph{timeliness}, \emph{relevance}, and \emph{utility}—to meet communication goals. Semantic approaches reduce the volume of generated, exchanged, and processed data while improving resource allocation\cite{kountouris2021semantics,uysal2022semantic}. This approach has been effectively applied in TN IoT networks and status-update systems\cite{yates2021age}. However, it requires greater attention in unified TN-NTN environments, where transmitting uninformative data leads to power waste and consequently shortens the lifetime of NTN platforms, exacerbating their stringent SWaP (Size, Weight, and Power) limitations.

Recent studies have initiated the incorporation of semantics in NTN networks \cite{ngo2024timeliness,liao2024information,liyanaarachchi20256g}. However, their primary focus has been limited to one aspect of semantics, i.e., \emph{timeliness}. In contrast, other aspects, such as the \emph{relevance} and \emph{utility} of information, which reflect the true significance of data, have been largely overlooked. Moreover, most of these studies focus on UAVs or single-satellite setups, disregarding the potential of networked satellite configurations and other platforms, such as high-altitude platforms (HAPs). By addressing these gaps and effectively integrating semantics-aware approaches within TN-NTNs, we can facilitate significant advancements for next-generation 6G networks.

To this end, in this article we propose integrating semantics-aware communication into a unified TN-NTN framework that includes a network of interconnected LEO satellites. In this approach, we utilize \emph{content-based} and \emph{query-based} semantic metrics, which capture not only the timeliness but also the relevance and utility of the information. Specifically, we employ the \emph{Query Version Age of Information (QVAoI)} metric for the first time in a multi-hop communication setup from a remote EH IoT device to a monitoring node, wherein LEO satellites serve as relaying nodes, and we investigate the optimal information handling policy for status updates within this network. The results indicate that, compared to the state-of-the-art (SoA) schemes, the optimization of appropriate semantics, especially QVAoI, leads to a significant reduction in the number of transmissions from the device, thereby decreasing dissemination in the LEO network. This reduction leads to lower energy consumption and cost savings by reducing the requirements for transmission, storage, and computation.

The remainder of this article is organized as follows: Section \ref{Sec_MainTechChallenges} outlines main technical challenges in unified 6G TN-NTNs; Section \ref{Sec_SemanticComm} introduces the semantics-aware communication approach; Section \ref{Sec_SemanticSystem} derives the optimal semantics-aware policy, evaluated in Section \ref{Sec_Results}; and Section \ref{Sec_Considerations} discusses implementation and scalability considerations. Finally, Section \ref{Sec_Conclusion} offers conclusions and future directions.

\section{Main Technical Challenges}

\label{Sec_MainTechChallenges}

Efficient information handling and data management are crucial in unified TN-NTNs due to inherent constraints such as high latency, mobility, interference, and energy limitations. Next-generation networks generate massive data volumes, making resource optimization essential for energy-efficient management and reliable service continuity across the 3D network. Key technical challenges in data management include:

\begin{itemize}[leftmargin=1em]
\item{\textbf{Latency Management:}} Unified TN-NTNs, particularly in satellite communications, can experience significant latency due to long signal travel distances. For instance, GEO satellites at an altitude of $35{,}786$ km introduce roughly $250$ ms two-way delay, adversely affecting real-time or time-critical applications such as voice calls, remote control, and monitoring systems. Multi-hop satellite links further amplify this issue, making timely and relevant information crucial.

\item{\textbf{Energy Consumption and Costs:}} Signals in TN-NTNs suffer substantial path loss from long-distance propagation and atmospheric attenuation, requiring higher transmission power and increasing energy consumption. This is particularly critical for power-limited devices such as IoT nodes, satellites, and UAVs. Moreover, UAVs and HAPs have limited energy, and continuous high-volume transmissions rapidly deplete their batteries, requiring frequent landings for replacement or recharging. Minimizing transmissions is thus essential to conserve energy, reduce maintenance, support sustainable communication, and enhance system lifecycle, thereby lowering operating expenses (OPEX) and total cost of ownership (TCO).

\item{\textbf{Handover and Service Continuity:}} Rapid movement of NTN nodes, such as LEO satellites, necessitates frequent handovers and routing adjustments. Continuous information and control exchanges strain network resources, so minimizing unnecessary transmissions is critical to ensure seamless connectivity.

\item{\textbf{Spectrum Scarcity and Interference:}} TN-NTNs operate in congested spectrum bands where coexisting transmissions cause interference, further aggravated by Doppler shifts from moving NTN nodes. Excessive data transmissions reduce spectral efficiency, making the minimization of redundant transmissions essential to mitigate interference and optimize spectrum use.

\item{\textbf{Massive and Diverse Data Traffic:}} TN-NTNs handle heterogeneous data types—real-time video, IoT telemetry, emergency alerts, and bulk downloads—each with distinct QoS requirements. Ineffective management of uninformative data can congest the network and degrade QoS for critical applications, including PPDR and telemedicine.

\item{\textbf{Caching and Real-Time Processing:}} NTNs and edge IoT devices have limited memory, storage, and processing capabilities, challenging large-scale real-time data communication. Overloading these resources with stale or uninformative data leads to processing delays and service degradation.
\end{itemize}

\section{Semantics-Aware Communication}
\label{Sec_SemanticComm}
Addressing the aforementioned technical challenges necessitates a careful \textit{focus on network and information management approaches}, considering the \textit{context and content of information within the entire chain of information handling process from the information generation to its utilization}. The semantics-aware approach achieves this by enabling goal-oriented configurations and optimizing the relevant semantics of information within the 3D network. It aims to significantly reduce—or even eliminate—the transmission of stale, uninformative, or unusable data while satisfying QoS requirements. As a result, substantial energy savings, latency reduction, and decreased data congestion can be realized in the unified network. This section first elaborates on the semantic attributes of information and associated metrics, then presents a semantics-aware status update setup as a core configuration for communication in unified TN-NTNs.

\subsection{Semantic Attributes}
The incorporation of semantics of information into communication networks has demonstrated substantial benefits, beginning with the introduction of the Age of Information (AoI)\cite{kaul2012real}, which captures the timeliness of information, and extending to other attributes, such as relevance and utility. The key semantic attributes of information are explained below.

\begin{itemize}[leftmargin=1em]
\item{\emph{Freshness or timeliness:}} AoI, going beyond the widely used metrics of delay and throughput, effectively represents the timeliness (or staleness) of data in the network since its generation. AoI considers the information chain from generation to reception, and optimizing it reduces unnecessary outdated data exchanges while freeing up resources for more efficient information handling.  
\item{\emph{Relevance:}} When sampling or generating data at the source, not all generated data conveys the correct piece of information content-wise. For instance, if the information content remains unchanged over time, generating or transmitting repetitive data to enhance freshness may be wasteful. This highlights a critical aspect of semantic metrics: the relevance of information, which emphasizes communicating data that carries relevant content at the source.
\item{\emph{Utility or value:}} Another shortcoming of AoI is that it primarily considers the chain of information from its generation to its reception, while neglecting its utilization at the destination node. For example, if the receiver is unavailable or too busy to process the received data, maintaining the timeliness and relevance of the information is insufficient. This introduces the third semantic attribute, i.e., utility or value of information, at the destination, emphasizing that information must be communicated when it is actually useful and valuable for utilization. 
\end{itemize}

These three attributes – timeliness, relevance, and utility – represent the important dimensions of semantics-aware communication, as they collectively capture the significance and value of data, encompassing its generation, transmission, and eventual utilization within the network.

\subsection{Semantic Metrics}
 The aforementioned semantic attributes are quantified through metrics, such as AoI, Version AoI (VAoI)\cite{yates2021vage}, Query AoI (QAoI)\cite{chiariotti2022query}, and Query VAoI (QVAoI)\cite{delfani2024semantics}. Optimizing these metrics in communication networks can significantly reduce resource usage while preserving the integrity of the conveyed information. 

\begin{itemize}[leftmargin=1em]
    \item \textbf{AoI} is a \emph{timeliness} metric that quantifies the time elapsed since the generation of the last successfully received data at the destination node. 
    \item \textbf{VAoI} is a semantic metric that jointly quantifies the \emph{timeliness} and \emph{relevance} of information by measuring the number of versions by which the receiver lags behind the source. A version of information is defined as any marked change in the content of information at the source. This metric has been analyzed and optimized in interconnected gossip networks\cite{kaswan2023age}.
    \item \textbf{QAoI} is another performance metric that represents both the \emph{timeliness} and \emph{utility} of information. QAoI considers the AoI only when there is a request from the destination node, i.e., only during query instances when the information is assumed to be useful to the receiver. 
    \item \textbf{QVAoI}, as an extension of both VAoI and QAoI, takes into account both content changes at the source and queries from the destination. This metric represents updates that are both \emph{relevant} and \emph{valuable}, alongside the \emph{freshest} ones. QVAoI is defined as the number of versions by which the receiver lags behind the source at query instances. This metric ensures that the system is not penalized in the absence of requests, and when the transmission of updates holds no value or utility for the receiver.
\end{itemize}

    \begin{figure*}[!t]
		\centering
		\begin{minipage}[b]{0.57\linewidth}
            \centering
	        \includegraphics[width=\linewidth]{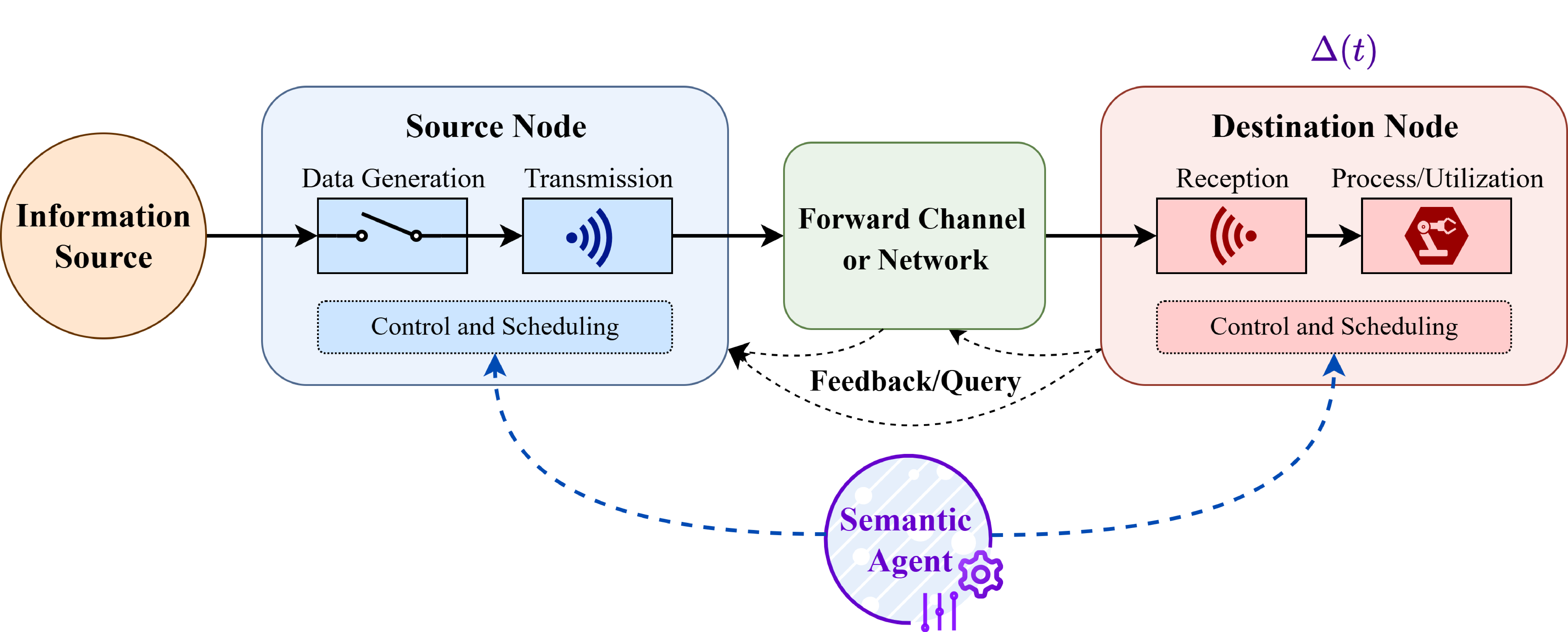}
            \caption{An end-to-end status update system equipped with a semantic agent that considers the generation, transmission, and utilization of information. The semantic metric $\Delta(t)$ quantifies the timeliness, relevance, and utility of information at the destination node.}
	        \label{fig_StatusUpdateSys}
		\end{minipage}
		\hfill
		\begin{minipage}[b]{0.40\linewidth}
            \centering
	        \includegraphics[width=\linewidth]{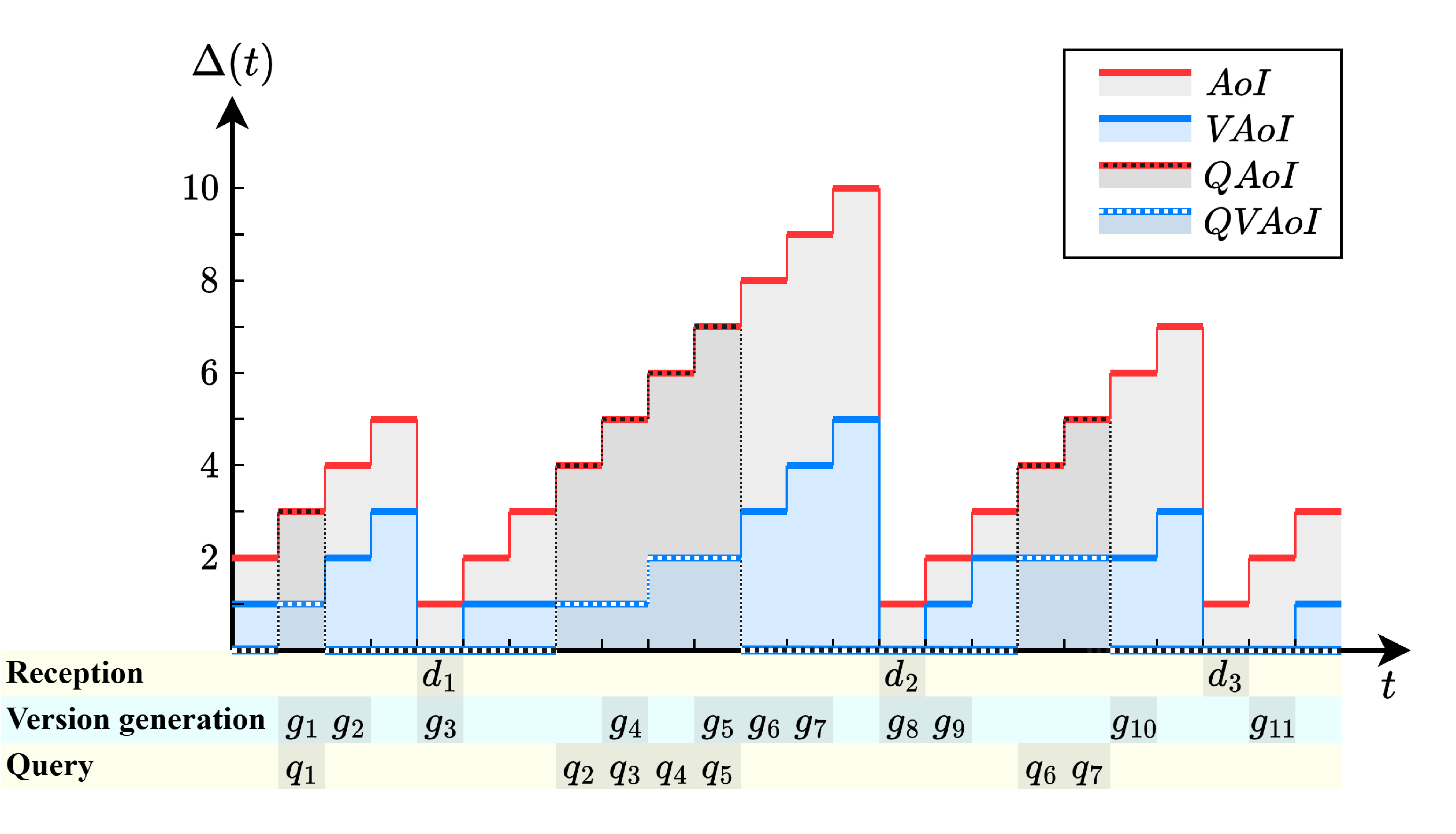}
	        \caption{Semantic metrics: AoI captures the timeliness of information, while VAoI and QAoI further encompass its relevance and utility. QVAoI integrates all three semantic attributes.} 
	        \label{fig_SemanticMetrics}
		\end{minipage}
        \vspace{-14pt}
	\end{figure*}

\subsection{Semantics-aware Scheduling}
In Fig. \ref{fig_StatusUpdateSys}, a basic end-to-end (E2E) status update system is shown, serving as a core configuration for data communication in IoT networks and real-time monitoring systems. The status of an information source is measured by a device acting as the source node and transmitted as \emph{update packets} to a destination node through a network or direct forwarding channel, where the packets are subsequently processed or utilized. In addition to the forward path, a backward channel may be present to transmit acknowledgment (ACK) feedback or queries. Such queries can trigger a \emph{pull-based} status update, in which the destination or other network nodes request the update packet from the source. In contrast, a \emph{push-based} setup forwards updates to the destination regardless of queries.

In this status update setup, the semantic metrics can be represented as a function of discrete time slots, $\Delta(t)$ for $t=0,1,2,\dots$. An illustration is provided in Fig. \ref{fig_SemanticMetrics}, where the data is generated in each time slot. We assume that the service time required for a generated update packet to reach the destination node is one time slot. Let $d_i$, $g_i$, and $q_i$ denote the reception, version generation, and query instances, respectively, for $i = 1,2,\dots$. At time $d_1$, the update packet generated in the previous slot is delivered, reducing the AoI to $1$ and the VAoI to $0$, as the source and destination now share the same version. Both QAoI and QVAoI are zero at this point, as no query has yet been received. Subsequently, AoI increases linearly until the next reception ($d_2$). Meanwhile, five versions are generated at the source ($g_3$ to $g_7$), raising VAoI to $5$, as the destination lags behind by five versions. During this period, four successive queries ($q_2$ to $q_5$) arrive, setting QAoI and QVAoI equal to AoI and VAoI, respectively. The evolution of the semantic metrics proceeds similarly thereafter.

In this E2E setup, a semantics-aware agent is employed to control the scheduling of the information handling chain, which encompasses the generation, transmission, and utilization of data, while accounting for the semantics of the information. Depending on the configuration, this agent can be positioned at the source, the destination, or both.  
The semantic agent determines the scheduling policy according to which update packets are communicated within the system. This policy optimizes the semantic metrics, aiming to deliver timely and informative data while ensuring the efficient utilization of network resources.

\section{Semantics-aware Unified TN-NTNs: \\ System Model and Optimization}
\label{Sec_SemanticSystem}

\subsection{System Model}
\begin{figure}[tb!]
	\centering
	\includegraphics[width=2.8in]{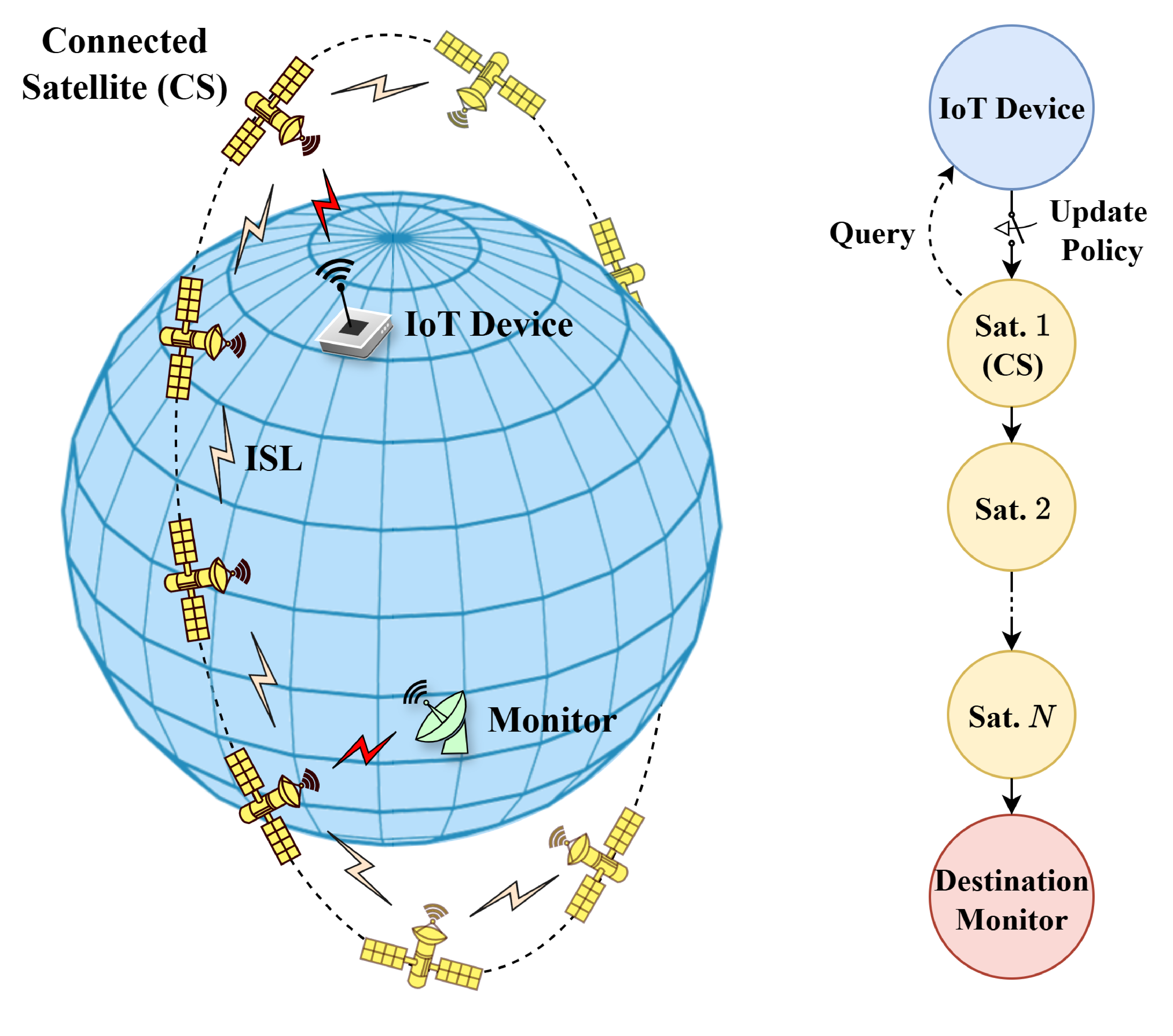}
	\caption{Status updates via the LEO satellite network: A remote IoT device equipped with a semantic agent transmits update packets to a monitoring node through a Connected Satellite (CS) and relay satellites. The agent schedules transmissions to optimize semantic metrics at the destination.}
	\label{fig_SatNet}
\end{figure}

Consider an IoT device (user equipment) located on a remote location on Earth, aboard an autonomous ship, or even on a HAPS, that measures and sends status updates (user data) from an information source—such as the state of an oil pipeline, temperature, or humidity—to a destination monitoring node where no direct terrestrial link exists. This scenario applies, for example, to environmental monitoring of distant locations (e.g., polar ice melting, air quality, temperature, pollution) or to PPDR use cases where terrestrial infrastructure has become unavailable. A connected LEO satellite can receive the device’s update packets and relay them to the monitoring node, either directly or via neighboring satellites in the same or adjacent orbital planes\footnote{We consider this connection is established within a visibility window during which the device remains connected to the same satellite without handovers.}. Communication between LEO satellites occurs through Inter-Satellite Links (ISLs), such as high-capacity optical ISLs. Fig. \ref{fig_SatNet} illustrates this scenario.

The objective is to determine an optimal update policy for transmitting updates from the device to the Connected Satellite (CS) by incorporating a semantic agent on the device that optimizes semantic metrics at the destination monitor. We consider a slotted time horizon in which the semantic agent sequentially decides whether to transmit or remain idle in each slot. We assume that the transmission of each packet from one node to another occupies a single time slot. The system model is described in detail below.

\textit{Unified network configuration:} The considered setup consists of an IoT device, $N$ LEO satellites serving as relay nodes, and a destination monitor configured as a line network (Fig. \ref{fig_SatNet}). The complete delivery of an update to the ground monitor involves a transmission from the device to the CS, multiple disseminations via ISLs within the LEO satellites, and a final transmission from the last satellite to the ground monitor.

\textit{Communication channels:} We assume an unreliable forward (uplink) channel with a defined success probability from the device to the CS, over which data packets are transmitted according to the update policy in each time slot (thus determining whether retransmission is required). An error-free backward (downlink) channel from the CS to the device\footnote{The satellite can employ a larger antenna and higher transmission power than the IoT device, which is why downlink communication is considered reliable, whereas uplink communication is regarded as unreliable.} enables the transmission of queries and ACKs confirming successful packet reception. Queries, generated by the CS, reflect the utility of the data in the network and may be scheduled according to deterministic policies, stochastic models based on network or monitor utilization, or the CS’s availability. For example, queries may be sent only when the CS can receive and forward update packets; otherwise, if it is busy or unable to forward the packets, they are withheld. The CS and other relaying satellites are assumed to forward each received update to the next node in the line network over reliable links\footnote{For unreliable dissemination links, feedback mechanisms and retransmissions at the relaying nodes should be considered. These aspects are omitted, as their detailed analytical modeling lies beyond the scope of this article.}.

\textit{Energy model of device:} The IoT device is assumed to have a rechargeable battery charged by harvesting ambient energy, such as solar or wind, reflecting common energy constraints in real-world scenarios. When the battery is depleted, the device cannot transmit updates. Otherwise, the semantic agent sequentially decides the \emph{action} for transmissions over a slotted time axis. Each battery unit is probabilistically charged at each time slot and normalized to the energy required for a single transmission; each transmission consumes one battery unit. 

\textit{Timestamps and versions:}
Each update packet from the source is assigned a timestamp and version number. New versions are generated at a specific rate or probability in each time slot. Upon receiving ACK feedback, the device learns the most recent timestamp and version stored at the CS and computes the semantic metrics accordingly.

In this article, version generations, query arrivals, and energy arrivals are modeled as stochastic Bernoulli processes, which provide a simple and effective representation of their \emph{rates}, defined as the average probability of occurrence per time slot\footnote{The Bernoulli assumption is justified by the long-horizon setting, where a large number of time slots ensures, via the law of large numbers, that these processes capture average behavior while keeping the model tractable.}. That is, each event occurs stochastically with a fixed average rate: new information versions are generated at the source according to an \emph{average version generation rate}; queries are sent by the CS (e.g., when it can forward update packets) with an \emph{average query rate}; and energy is harvested by the device’s battery with an \emph{average charging rate}.

\textit{Remark:} In the unified network, the semantic metrics at the destination monitor are directly determined by the semantic metrics at the CS.
Therefore, the optimization of semantic metrics at the destination monitor reduces to their optimization at the CS.\footnote{Particularly, in scenarios with reliable dissemination, the AoI (QAoI) at the destination monitor equals the AoI (QAoI) at the CS plus $N$, due to the delay from the $N$ relaying nodes, each requiring one time slot to deliver a packet. Similarly, the VAoI (QVAoI) at the destination monitor equals that at the CS plus a Binomial random variable representing the number of version updates generated at the source over $N$ slots, since each version at the CS reaches the monitor after $N$ slots while the source independently generates new versions according to a Bernoulli process in each slot.}
Moreover, minimizing the transmission of a single update from the device to the CS directly reduces the propagation of $N$ subsequent disseminations throughout the unified network. This indicates that both the energy consumption at the IoT device and the number of transmissions in the network are influenced by the scheduling policy in the same proportion, highlighting the importance of efficient information handling.

By deploying a semantic agent on the device for optimal transmission scheduling, the configuration in Fig. \ref{fig_StatusUpdateSys} is obtained, where the IoT device serves as the source node and the CS as the immediate destination.

\subsection{Optimal Information Handling Policy}

The optimization of semantic metrics within this framework can be formulated as a Markov Decision Process (MDP). The objective is to minimize the long-term expected value of each metric (AoI, VAoI, QAoI, or QVAoI) by deriving an optimal update policy. At each time slot, this policy dictates whether to transmit or withhold an update based on the current system state, defined by the semantic metric’s instantaneous value, the device’s battery status, and the query issued by the CS. The MDP explicitly incorporates statistical descriptions of the underlying stochastic processes, including energy arrivals, query arrivals, and information source version generation, each of which is modeled as an independent Bernoulli random variable with its specified rate or average occurrence probability per time slot. It can be shown that the state space of these MDPs is (weakly) communicating, ensuring the existence of an optimal policy, which can be derived via the Bellman equations and the Relative Value Iteration Algorithm (RVIA).

\emph{By optimizing semantic metrics in the system, an optimal scheduling policy is derived, which offers benefits from two perspectives: first, the freshest and most informative status updates are transmitted within the network, enabling precise monitoring and informed decision-making at the monitor, all while adhering to energy constraints; second, the number of transmissions from the device to the CS, and consequently the number of relayed packets among the satellites, is reduced by excluding stale and uninformative data from the network. This leads to a significant reduction in energy and cost consumption within the network}, as will be demonstrated in the next section.

\section{Evaluation Results}
\label{Sec_Results}
In this section, we study the performance of the described system in delivering timely, relevant, and useful status updates to the network. This is captured by QVAoI at the CS, which directly determines the QVAoI at the satellite nodes and the destination monitor. The following policies adopted by the semantic agent are compared in terms of their ability to provide the optimal QVAoI in the network: 1) \textbf{Greedy policy}: transmit an update as soon as energy becomes available and the battery is not empty; 2) \textbf{AoI-aware}, 3) \textbf{VAoI-aware}, 4) \textbf{QAoI-aware}, and 5) \textbf{QVAoI-aware policies}: transmit the updates when optimizing the time average of AoI, VAoI, QAoI, and QVAoI, respectively. 
These optimal policies are obtained by solving the corresponding MDP problems using RVIA. 

In the following simulations, expected values are computed using Monte Carlo methods in MATLAB. We evaluate the performance of the aforementioned policies under various system parameters, specifically low and moderate energy charging and query rates (as specified in the following subsections), by fixing the channel success probability to $80\%$ and the version generation rate to $25\%$.

\begin{figure}[t!]
	\centering
    \includegraphics[trim={1.5cm 0 0 0},clip,width=0.96\linewidth]{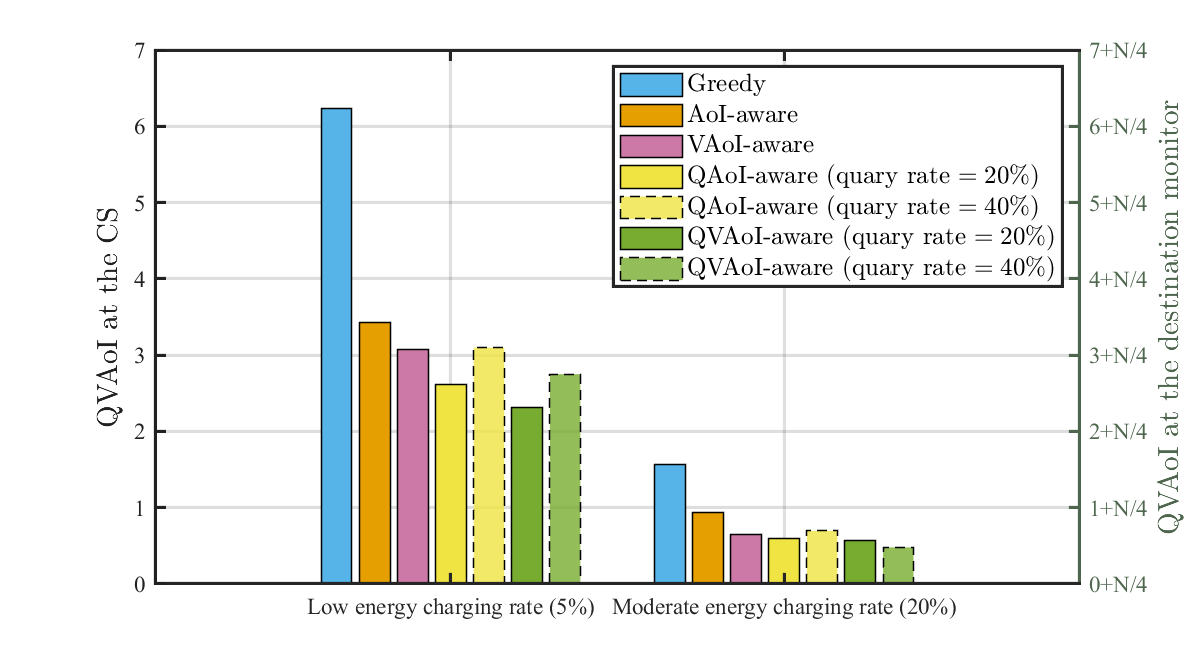}
    \caption{QVAoI at the CS (left $y$-axis) and at the destination monitor (right $y$-axis), $N$ hops from the CS, for $20\%$ (solid bars) and $40\%$ (dashed bars) query rates: Incorporating additional semantic attributes yields fresher and more informative network data.}
    \label{fig_QVAoIq0.2q0.4}
\end{figure}

\subsection{Performance of Various Policies}
\label{sec_PerfAnalysis}
In Fig. \ref{fig_QVAoIq0.2q0.4}, the performance of the greedy and semantics-aware policies is evaluated for two query rates ($20\%$ and $40\%$) under low ($5\%$) and moderate ($20\%$) energy charging rates. In all cases, adopting the optimal semantics-aware policies substantially improves system performance relative to the greedy policy, delivering the freshest and most informative data from the information source to the CS (left $y$-axis), and consequently to the destination monitor (right $y$-axis), whose average QVAoI equals the average QVAoI at the CS plus the average number of version generations over $N$ relaying slots (i.e., the version generation rate multiplied by $N$, or $N/4$). The results also show that VAoI-aware and QVAoI-aware policies outperform AoI-aware and QAoI-aware policies, respectively. By accounting for relevance, timeliness, and utility, these policies provide a more effective scheduling scheme for transmitting updates when they are most informative, while using the same energy resources.

By increasing the query rate from $20\%$ to $40\%$ (shown with dashed bars), the performance of the semantics-aware system may either improve or decline, depending on the energy charging rate, while consistently outperforming the greedy policy. For moderate to high energy charging rates, the system becomes less energy-constrained, allowing for the transmission of more updates. Consequently, the performance of the system becomes more dependent on the utility of the information, i.e., the query rate; therefore, an increase in query rate enhances the performance in terms of QVAoI. However, when the energy charging rate is low, the system becomes energy-restricted, and exerting more pressure by sending additional queries leads to a greater number of query instances with the same amount of updates, resulting in a deterioration of the QVAoI.

\subsection{Optimal Actions}
\begin{figure}[t]
	\centering
	\includegraphics[width=3.1in]{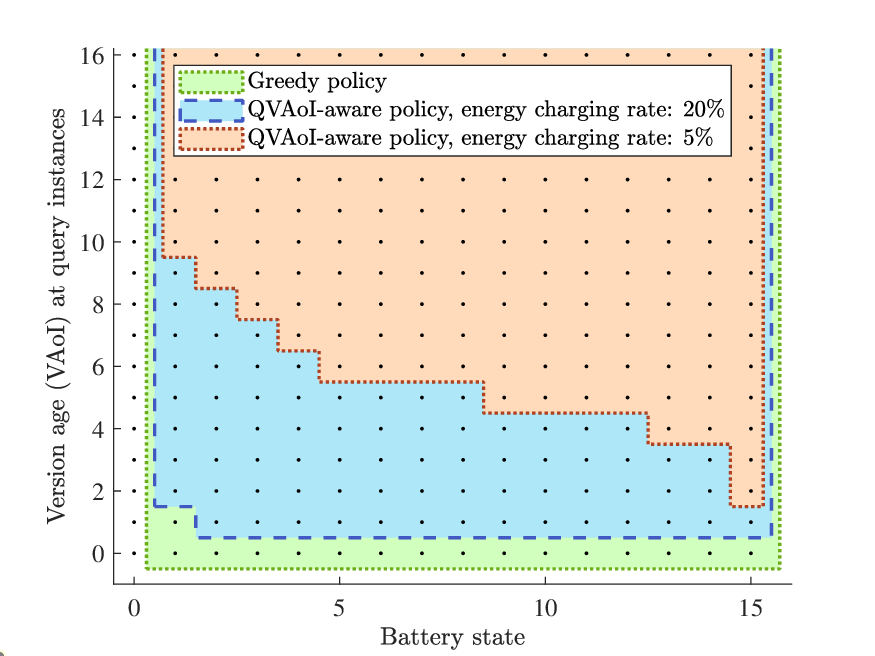}
	\caption{Transmission regions for the greedy and semantics-aware policies: The semantics-aware policy employs a threshold-based mechanism, triggering updates when the QVAoI exceeds the threshold. The optimal policy raises thresholds at lower energy charging rates, particularly under low battery states. By contrast, the greedy policy disregards QVAoI, transmitting solely when the battery is non-empty.}
	\label{fig_Actions}
\end{figure}

To further illustrate the policies, Fig. \ref{fig_Actions} presents the actions taken by the agent under both the greedy and the semantics-aware policies. The QVAoI-aware policy is used to represent the semantics-aware policies, as it captures all semantic attributes and achieves the best performance among the reported results. Two charging scenarios—low and moderate rates—are examined for the optimal QVAoI-aware policy, with a fixed query rate of $30\%$. 
The selected action (transmit or remain idle) depends on the VAoI at query instances (since no transmission occurs without a query) and on the battery state, i.e., the number of charged and usable battery units. In the figure, this produces transmission regions in which updates are sent; otherwise, the device stays idle. The greedy policy considers only the battery state and transmits whenever the battery is non-empty, yielding the largest region (shown in green). In contrast, the optimal semantics-aware policy incorporates the VAoI at query instances (or QVAoI) and operates as follows:

\begin{itemize}[leftmargin=1em]
    \item It never transmits updates when the VAoI is zero, i.e., when the device and the CS contain identical versions or content of information, as such updates are uninformative.
    \item It sets VAoI thresholds for each battery state to avoid transmitting updates when the CS versions are not sufficiently outdated. The optimal policy prioritizes transmissions at higher VAoI values, since sending at lower values depletes the battery. This can lead to long intervals without updates, during which the CS misses many newer versions, causing the QVAoI to grow substantially. As shown in Fig. \ref{fig_Actions}, the transmission region for the lower charging rate ($5\%$, light orange) is smaller and has higher VAoI thresholds than in the higher charging rate scenario ($20\%$, blue).
\end{itemize}

\subsection{Energy Consumption and Dissemination Reduction}
\begin{figure}[tb!]
	\centering
	\includegraphics[width=2.9in]{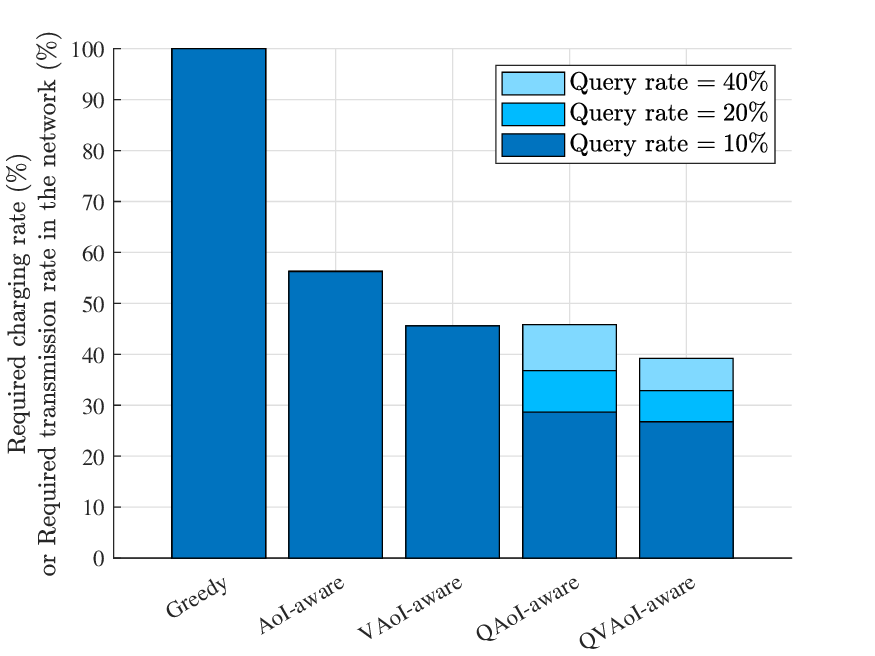}
	\caption{The required energy charging rate, or equivalently, the required number of transmissions under different policies (relative to the greedy policy), to maintain a specific QVAoI at various query rates.}
	\label{fig_NeededEnergy}
\end{figure}

The results obtained in Section \ref{sec_PerfAnalysis} indicate that to maintain a specific level of QVAoI in the system, different policies require different energy charging rates. Fig. \ref{fig_NeededEnergy} illustrates the required energy charging rate for various policies to achieve an average QVAoI of $1.5$. This quantity is presented relative to (i.e., divided by) the energy charging rate of the greedy policy. As observed, semantics-aware policies significantly improve energy efficiency, with the QVAoI-aware policy requiring only $27\%$ of the energy charging rate (i.e., $73\%$ less) compared to the greedy policy at a query rate of $10\%$. This substantial reduction in energy consumption is achieved by discarding stale, irrelevant, and non-utilizable information in the network. The reduced number of updates from the IoT device decreases the number of disseminations in the relaying satellites by the same ratio and also results in lower storage and processing requirements in the unified TN-NTN.

\emph{Query-Aware vs. Query-Agnostic:}
The query-aware policies (QAoI- and QVAoI-aware) restrict transmissions to query instances, providing updates only when the information can be utilized within the network. This approach is particularly effective in energy-constrained settings, where the energy charging rate is lower than the query rate. By contrast, when energy is abundant and the charging rate exceeds the query rate, transmissions may also occur outside query instances, enabling query-agnostic policies (AoI- or VAoI-aware) to achieve lower QVAoI at the cost of higher energy consumption. Thus, when the charging rate is lower than the query rate, query-aware policies outperform query-agnostic ones by either maintaining comparable QVAoI with reduced energy use or attaining better QVAoI at the same charging level. Conversely, when the charging rate exceeds the query rate, query-agnostic policies become more effective by leveraging surplus energy to transmit updates beyond query instances.

\section{Implementation and Scalability}
\label{Sec_Considerations}
Semantics-aware information handling represents a promising approach for optimizing the communication of timely, relevant, and useful data, thereby enabling more cost- and energy-efficient unified TN-NTNs. However, it also entails practical requirements that must be carefully addressed. We discuss the implementation and scalability of the proposed semantics-aware information handling scheme.

\subsection{Implementation of the Optimal Policy}
The quantification of semantic metrics requires timestamps for each update packet or version indexes for each generated version, and an ACK feedback mechanism from the CS to the device. Using these elements, the semantic agent on the device can compute the semantic metrics and, in accordance with the scheduling policy, determine the appropriate transmission actions. \emph{A key requirement in this process is the definition of what constitutes a new version}. This definition depends on changes in the content of the data, the question arises as to what degree of modification, minor or major, should be sufficient to establish a new version. This can be determined based on the specific requirements of the task at hand.

In addition, characterizing the optimal policy can impose a processing burden on the device, challenging low-energy systems. Nevertheless, the optimal semantics-aware policy from the infinite-horizon MDP yields a stationary look-up table of $0$ and $1$ elements ($0$ for idleness, $1$ for transmission), defined by the system state—specifically, battery level and VAoI at the CS. This policy can be computed offline\footnote{The optimal policy can be obtained on a resource other than the IoT device, after which the resulting policy can be sent to and loaded onto the device. The computational complexity of RVIA scales with the square of the state space, i.e., the square of the product of the battery size and the VAoI bound.} only once, as long as system parameters remain fixed. With the policy stored in a look-up table and the current state known, the optimal action for each time slot can be determined online. Moreover, when the policy has a threshold-based structure, storing the full table is unnecessary; it is sufficient to record only the threshold for each battery level at which the action switches from $0$ to $1$. This allows for a lightweight scheme suitable for low-power IoT devices.

\subsection{Scalability}
\emph{Multiple satellites:} The proposed semantics-aware approach focuses exclusively on the transmission link between the device and the CS, thereby directly influencing the device's energy consumption and indirectly affecting satellite dissemination through ISLs. This approach can therefore be easily scaled to accommodate a large number of satellites, $N$, as relay nodes, enabling support for long-distance communications\footnote{If queries are issued by the farther nodes rather than the CS, potential delays in receiving them must also be carefully considered.}.

\emph{Multiple devices:} The system model considered in Section \ref{Sec_SemanticSystem} focuses on an E2E configuration with a single source (device) and a single destination. This framework can be naturally extended to multiple devices and destinations, provided that each source–destination pair maintains a dedicated transmission route. However, when multiple devices share a common CS or relaying nodes, optimal scheduling requires additional considerations, such as avoiding contention and ensuring fairness among devices. Nonetheless, the proposed E2E approach offers a useful foundation for generalizing the model, since optimizing the transmissions of one device typically results in efficient resource utilization, allowing others to transmit their updates.

\section{Conclusion and Future Directions}
\label{Sec_Conclusion}
This article investigated semantics-aware communication in unified TN-NTNs and proposed an optimal policy for managing status updates from energy-constrained IoT devices to monitoring nodes via satellite networks. By evaluating various semantic metrics under different charging and query arrival scenarios, we showed that semantics-aware communication can substantially reduce network transmissions, especially when relevance and utility are considered alongside timeliness. This reduction lowers energy consumption and decreases charging demands by up to $73\%$ compared to state-of-the-art methods. These results highlight the potential of semantics-aware information handling to enable more sustainable and efficient next-generation TN-NTNs. Future work could extend this framework to multi-device and multi-agent scenarios, integrate aerial network components for greater adaptability, investigate energy-harvesting models with unknown statistics or imperfect feedback channels, and assess the impact of semantic agents on overall network energy efficiency.

\bibliographystyle{IEEEtran}
\bibliography{Refs}

\end{document}